\documentclass[aps,pra,twocolumn,groupedaddress,showpacs, floatfix]{revtex4-1}
\usepackage{amsmath}

\usepackage{graphicx}
\usepackage{bm}
\usepackage[mathlines]{lineno}

\usepackage{amsmath}
\usepackage{amssymb}
\usepackage{mathtools}
\usepackage[shortlabels]{enumitem}
\usepackage{gensymb}
\usepackage{color}
\usepackage{physics}

\begin{document}

\title{Circularizing Rydberg atoms with time-dependent optical traps}
\author{Ryan Cardman} \email{rcardman@umich.edu}
\author{Georg Raithel}
\affiliation{Department of Physics, University of Michigan, Ann Arbor, Michigan 48105, USA}

\date{\today}
\begin{abstract}
We discuss three proposed schemes of initializing circular-state Rydberg atoms via optical couplings provided by the ponderomotive effect in contrast to the current circularization methods that utilize electric-dipole interactions. In our first proposed method, a radial optical trap consisting of two Laguerre-Gaussian beams of opposite winding numbers transfers orbital angular momentum to the Rydberg atom, providing a first-order coherent coupling between an F-state and a circular state. Additionally, we propose a one-dimensional ponderomotive optical lattice modulated at rf frequencies, providing quadrupole-like couplings in the hydrogenic manifold for rapid adiabatic passage through a series of intermediate Rydberg states into the circular state. For the third proposed scheme, a two-dimensional ponderomotive optical lattice with a time-orbiting trap center induces effectively the same coupling as a $\sigma^{+}$ or $\sigma^{-}$-polarized rf field of tunable purity for all-optical rapid adiabatic passage into the circular state.
\end{abstract}
\pacs{32.80.Rm, 32.80.Qk, 32.80.Pj}

\maketitle
\section{Introduction}
Circular-state (CS) Rydberg atoms have maximum orbital angular momenta and reside in the extreme Zeeman sublevels. Electric-dipole selection rules permit spontaneous emission solely between adjacent CSs, extending lifetimes to the order of ms. Because of this feature allowing sufficient time for making spectroscopic measurements, CS Rydberg atoms are desirable for cavity QED  experiments~\cite{Haroche2013} and high-precision spectroscopy \cite{Ramos2017}. Examples of two ongoing experiments consist of a linear chain of trapped CS Rydberg atoms experiencing dipole-dipole and van der Waals interactions for quantum simulation~\cite{Nguyen2018} and a precise measurement of the Rydberg constant for solving the proton radius puzzle \cite{Pohl2010,Ramos2017}.

Because plane-wave electromagnetic fields change the internal angular momentum to an atom by only up to one $\hbar$ (in first order), one cannot use standard laser excitation to prepare a sample of CS Rydberg atoms, which usually have $\simeq20\hbar$ units of orbital angular momentum or more. Two popular methods of circularization are the crossed-fields method~\cite{Delande1988,Hare1988, Morgan2018} and the rapid adiabatic passage (RAP) method~\cite{Hulet1983,Nussenzveig193,Lutwak1997}. In the crossed-fields method, perpendicular electric and magnetic fields with slowly-varying amplitudes are applied to the system to adiabatically switch atoms in a low-$|m_{l}|$ state into the CS. The initial field magnitudes are chosen such that the Stark splitting is much larger than that of the Zeeman interaction. With this initial configuration, the outermost levels of the hydrogenic manifold are Stark states with $m_{l}\simeq0$ and are accessible by laser excitation. As the electric field is adiabatically switched off, while the transverse magnetic field remains fixed or adiabatically increases, the fields transfer the atom to the CS. While this is an effective method, it requires efficient suppression of electric-field noise. In the RAP method, linearly-polarized rf waves couple states with low magnetic quantum numbers $|m_{l}|\leq3$ to the CS. In this scheme, the electric and magnetic fields are parallel and lift the degeneracies of the hydrogenic states. For rf waves of a chosen frequency, the relevant dressed states nearly cross at a specified electric field. Electric-dipole coupling induced by an rf field turns this crossing into a multi-level avoided crossing, which permits adiabatic switching of the atoms from a low-angular-momentum state, accessible by lasers, to the CS via scanning of the electric field. Applications which require parallel electric and magnetic fields or a quantization axis defined by the Stark interaction find the RAP method more favorable than the crossed-fields method~\cite{Ramos2017,Lutwak1997}, for the latter would require diabatic switching of the atoms into the Stark-dominated regime by a sudden turn-on of an electric field parallel to the magnetic field subsequent to the circularization~\cite{Lutwak1997}.

The aforementioned methods employ slowly varying perturbations to the atomic system for efficient circularization; however, there has been recent interest in fast transitions into the circular state with purely $\sigma^{+}$-polarized rf fields~\cite{Signoles2017}. Simulations based on quantum optimal control theory can be performed to choose rf fields with appropriate relative phases and amplitudes in order to optimize the speed of circularization~\cite{Patsch2018}. In such methods, multiple hydrogenic states are excited at once, making the process analogous to a transition from one coherent state to another.

The quantum dynamics of the methods discussed arise from the term proportional to $\textbf{A}\cdot\textbf{p}$ in the minimal coupling Hamiltonian for a charged particle. This term describes electric-multipole transitions and the Stark effect. The term proportional to $A^{2}$ describes the diamagnetic and ponderomotive shifts (e.g., the Kapitza-Dirac effect in electrons~\cite{Bucksbaum1988,Freimund2001}). In Rydberg atoms with weakly-bound valence electrons, a rapidly oscillating electric field pushes the Rydberg electron to regions of intensity minima by means of the ponderomotive interaction~\cite{Dutta2000}, thereby exerting a net force on the entire atom. Therefore, this term becomes significant when dealing with Rydberg atoms. In this paper, we discuss methods of circularization involving ponderomotive interactions of the Rydberg electron, which are due to the $A^{2}$ part of the minimal coupling Hamiltonian.

Hermite-Gaussian (HG) modes of electromagnetic waves contain single units of angular momentum, whereas a properly prepared Laguerre-Gaussian (LG) mode of winding number $m$ has an angular momentum of $m\hbar$ per photon that can be on the order of a CS Rydberg atom's angular momentum. In section 2 of this paper, we propose a method of coherent Rydberg atom circularization with two co-propagating LG beams of winding numbers with the same magnitude $|m|$ but opposite signs. If the wavelengths of the beams are chosen appropriately, inelastic, coherent scattering between the modes, effected by the ponderomotive interaction, enables direct coupling of a low-$m_{l}$ Rydberg level with the CS. Multipole interactions of LG modes and Rydberg atoms have been discussed, where matrix elements coupling low-angular momentum states to high-angular momentum hydrogenic states were calculated~\cite{Mukherjee2018,Rodrigues2016}, but these methods differ from ours, as they involve electric-multipole transitions due to the $\textbf{A}\cdot\textbf{p}$ term, not ponderomotive interactions.

In section 3, we discuss an optically-based RAP scheme that involves electric-quadrupole-equivalent coupling between Stark states of different $m_{l}$ by means of an rf-modulated ponderomotive optical lattice (POL)~\cite{Knuffman2007,Moore2015}. This proposed scheme also allows the atoms to remain trapped during the circularization, does not require rf fields in the atom-field interaction region, and enables circularization of atoms with a spatial selectivity on the order of $\mu$m.

In section 4, we discuss atoms in a two-dimensional POL with its trap center modulated in a circular motion at rf frequencies, analogous to the TOP trap used for Bose-Einstein condensation~\cite{Petrich1995}. The resulting ponderomotive interaction has the same effect as electric-dipole couplings by a purely $\sigma^{+}$ or $\sigma^{-}$-polarized rf field. This time-orbiting ponderomotive optical lattice (TOPOL) improves the efficiency of the RAP scheme by preventing ``leakage" transitions~\cite{Nussenzveig193, Hulet1983, Lutwak1997}.

In section 5, we discuss the advantages and disadvantages of our proposed all-optical schemes in comparison to the previously proposed and demonstrated non-optical methods. 
\section{Circularization by Laguerre-Gaussian laser modes}
\subsection{Theoretical model}
\begin{figure}[h!]
\begin{center}
\includegraphics[scale=0.25]{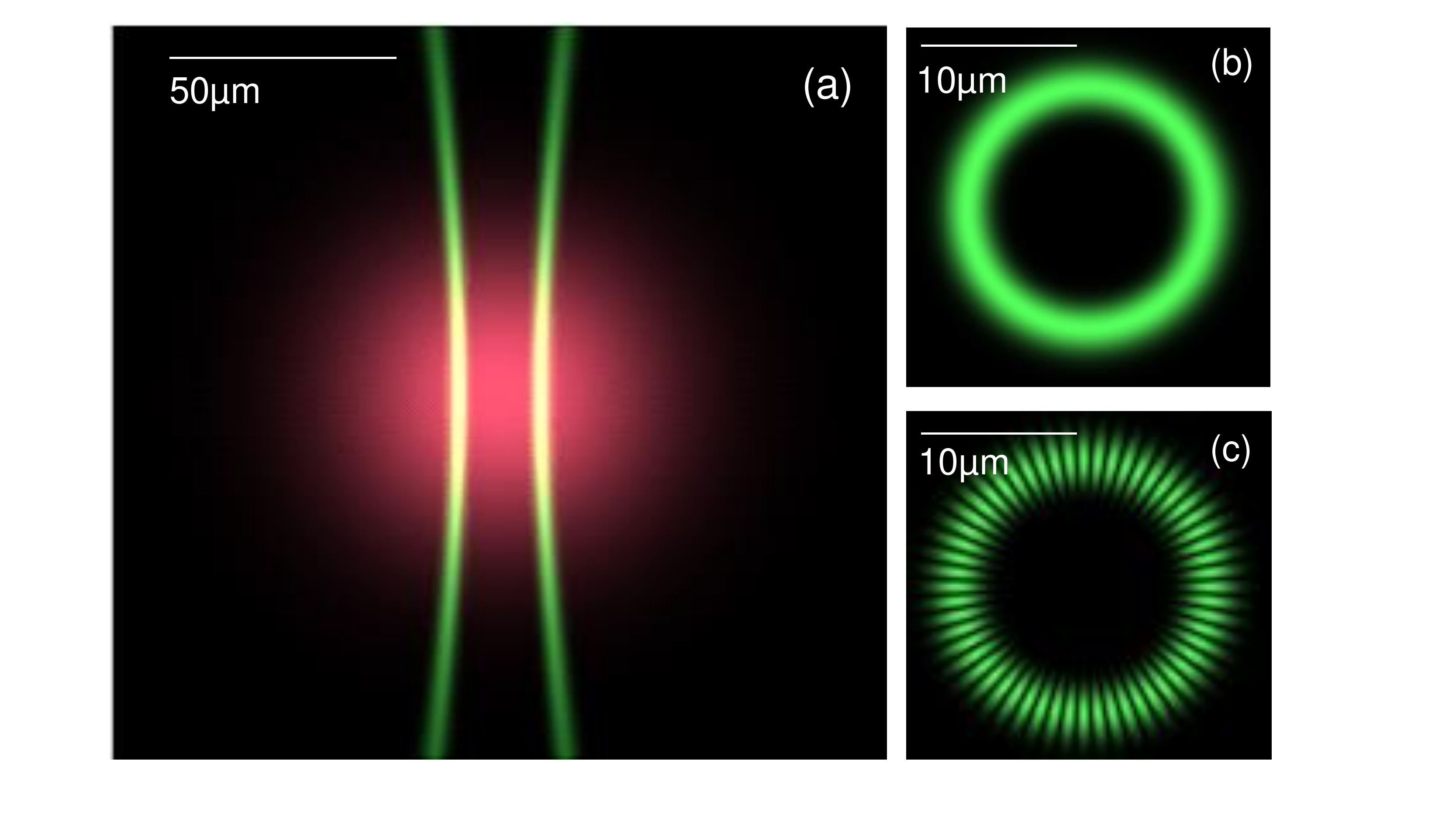}
\caption{For all figures above, $m=14$. In (a), the longitudinal intensity profile of the two superimposed beams with $\lambda_{1}=536$ nm and $\lambda_{2}=532$ nm is shown as they overlap a cloud of ultracold $^{85}$Rb. The diameter of a $21$F Rydberg atom and $n=32$ CS are at the order of a hundredth of the beam's diameter at $z=0$. In (b), the time-independent part of the ponderomotive potential $V_{p}(\textbf{r})$ is plotted at $z=0$; in (c), we show the magnitude of the time-dependent part of the ponderomotive potential $V_{C}$ on the xy-plane for $t=0$.}
\label{fig:potentials}
\end{center}
\end{figure}

Consider a coherent superposition of two LG modes (1 and 2), with respective field amplitudes $\mathcal{E}_{1}$ and $\mathcal{E}_{2}$, respective winding numbers $-m$ and $m$, and respective angular frequencies $\omega_{L_{1}}$ and $\omega_{L_{2}}$. The longitudinal intensity profile of this superposition is shown in Fig.~\ref{fig:potentials}(a) for the case of $m=14$. For a mode with a radial index $p=0$, we see that the optical field~\cite{Allen1992} can be described by
\begin{multline}
\label{eq:LGField}
   \textbf{E}(\textbf{r},t)=\frac{\hat{\epsilon}\mathcal{E}_{1}w_{0_{1}}}{2w_{1}(z)}\sqrt{\frac{1}{|m|!}}\bigg[\frac{\sqrt{2}\rho}{w_{1}(z)}\bigg]^{|m|} L^{|m|}_{0}\bigg[\frac{2\rho^{2}}{w_{1}(z)^{2}}\bigg]\\\times\exp\bigg[\frac{-\rho^{2}}{w_{1}(z)^{2}}\bigg]\exp\bigg[\frac{-ik_{1}\rho^{2}}{2R_{1}(z)}\bigg]\exp[i\psi_{1}(z)]\\\times\exp[i(m\phi-k_{1}z)+i\omega_{L_{1}}t]\\+\frac{\hat{\epsilon}\mathcal{E}_{2}w_{0_{2}}}{2w_{2}(z)}\sqrt{\frac{1}{|m|!}}\bigg[\frac{\sqrt{2}\rho}{w_{2}(z)}\bigg]^{|m|}L^{|m|}_{0}\bigg[\frac{2\rho^{2}}{w_{2}(z)^{2}}\bigg]\\\times\exp\bigg[\frac{-\rho^{2}}{w_{2}(z)^{2}}\bigg]\exp\bigg[\frac{-ik_{2}\rho^{2}}{2R_{2}(z)}\bigg]\exp[i\psi_{2}(z)]\\\times\exp[-i(k_{2}z+m\phi)+i\omega_{L_{2}}t]+\text{c.c.},
\end{multline}
where
\begin{align}
    w_{i}(z)&=w_{0_{i}}\sqrt{1+(z/z_{R_{i}})^{2}},\\
    R_{i}(z)&=z+\frac{z^{2}_{R_{i}}}{z},\\
    \psi_{i}(z)&=(|m|+1)\arctan{\big(z/z_{R_{i}}\big)},
\end{align}
and $\textbf{r}=(x,y,z)=(\rho,\phi,z)$ is a position vector in the laboratory frame. The vector $\textbf{r}$ is the vectorial sum of the atom's center-of-mass position $\textbf{R}$ and the relative coordinate $\textbf{r}_{e}$ of the Rydberg electron. We assume that the beam has a linear polarization described by the unit vector $\hat{\epsilon}$. The parameters  $z_{R_{i}}$ and $w_{0_{i}}$ are the Rayleigh ranges and waists of the beams, respectively for $i=1$ and 2. Note that we use the convention $\mathcal{E}_{i}=\sqrt{\frac{2I_{i}}{c\epsilon_{0}}}$ and $I_{i}=\frac{2P_{0_i}}{\pi w_{0_i}^{2}}$ with a power $P_{0_i}$ and a peak intensity $I_{i}$ for mode $i$. As the energy splittings from a low-$m_{l}$ Rydberg state to a CS Rydberg level range from the order of $h\times$~GHz to $h\times$~THz, the wavelengths of the two modes need only differ by a few nm or less. Thus, it is reasonable to focus the co-propagating beams with the same optics and assume similar Rayleigh ranges.

Electromagnetic fields can be introduced in the minimal coupling Hamiltonian by including the vector potential $\textbf{A}(\textbf{r},t)$ and scalar potential $\Phi(\textbf{r})$. For $\Phi=0$, noting that $\textbf{E}=-\partial_{t}\textbf{A}$ and using the harmonic nature of the assumed fields, we can include the vector potential $\textbf{A}$ operator in the Hamiltonian $H$.

After including the harmonic vector potential, we see that the minimal coupling Hamiltonian becomes
\begin{equation}
\label{eq:ham}
    H=\frac{1}{2m_{e}}(\textbf{p}+e\textbf{A})^{2}.
\end{equation}
The term proportional to $A^{2}$ consists of a time-dependent potential $V_{C}(\textbf{r},t)$ coupling a low-angular momentum Rydberg level to the CS and a time-independent electron trapping potential $V_{p}(\textbf{r})$. Both of these potentials seen by a Rydberg electron on the $xy$-plane are shown in Figs. ~\ref{fig:potentials}(b)~and~\ref{fig:potentials}(c). A scattering interaction between the two LG photons with opposite $m$ is responsible for the time-dependent part of the ponderomotive potential. This is given by
\begin{multline}
\label{eq:circ}
  V_{C}(\textbf{r},t)=4^{|m|}\frac{(2|m|)!}{|m|!}\sqrt{\frac{4\pi}{(4|m|+1)
    !}}\bigg(\frac{e^{2}\sqrt{I_{1}I_{2}}w_{0_{1}}w_{{0}_{2}}}{ m_{e} c\epsilon_{0}\omega_{L_{1}}\omega_{L_{2}}}\bigg)\\ \times r^{2|m|}[w_{1}(r\cos{\theta})w_{2}(r\cos{\theta})]^{-(|m|+1)}\\
   \times\exp(-r^{2}\sin^{2}{\theta}[w_{1}(r\cos{\theta})^{-2}+ w_{2}(r\cos{\theta})^{-2}])\\
    \times \big[Y^{2m}_{2|m|}(\theta,\phi)S(\textbf{r})\exp(-i\omega_{b}t)\\
    +Y^{-2m}_{2|m|}(\theta,\phi)S^{*}(\textbf{r})\exp(i\omega_{b}t)\big],
\end{multline}
with a phase term
\begin{multline}
S(\textbf{r})=\exp\bigg[\frac{i r^{2}\sin^{2}{\theta}}{2}\bigg(\frac{k_{2}}{R_{2}(r\cos{\theta})}\\
-\frac{k_{1}}{R_{1}(r\cos{\theta})}\bigg)\bigg]\exp\{i[\psi_{1}(r\cos{\theta})-\psi_{2}(r\cos{\theta})]\}\\
\times\exp[i(k_{2}-k_{1})r\cos{\theta}],
\end{multline}
where $\omega_{b}=\omega_{L_{2}}-\omega_{L_{1}}$. From Eq.~\ref{eq:circ}, it is seen that the time-dependent potential $V_{C}$ drives transitions between Rydberg-state pairs with energy difference $\hbar\omega_{b}$. The spatial structure of $V_{C}$, visualized in Fig.~\ref{fig:potentials}(c), determines the coupling matrix element between them.  For the conditions discussed in this paper, $S(\textbf{r})$ is virtually identical to unity; therefore, we set $S(\textbf{r}) = 1$.

\begin{figure}[h!]
\begin{center}
\includegraphics[scale=0.7]{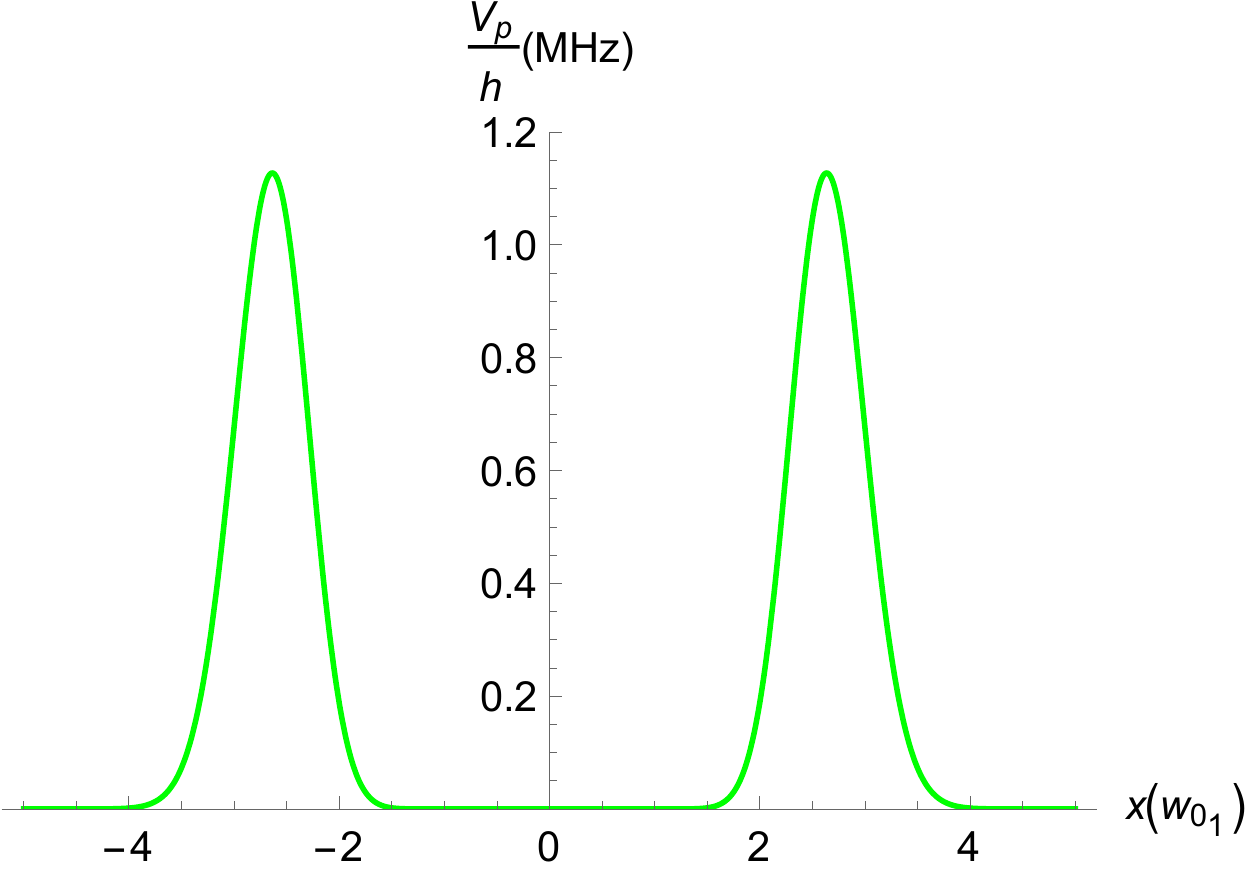}\\
\caption{Ponderomotive electron trapping potential included in the Hamiltonian as seen along the $x$-axis of the laboratory frame for $\lambda_{1}=536$ nm and $\lambda_{2}=532$ nm, where the origin is placed at the center of the LG modes of $m=14$ with $w_{0_{1}}=3.41\text{ } \mu$m and $w_{0_{2}}=3.39\text{ } \mu$m. This potential is plotted for total powers $P_{0_1}=P_{0_2}=150.0$ mW. }
\label{fig:trapping_potential}
\end{center}
\end{figure}

The static electron trapping potential $V_{p}$ is given by
\begin{multline}
V_{p}(\textbf{r})=\frac{e^{2}I_{1}w^{2}_{0_{1}}}{2 |m|!m_{e} c\epsilon_{0}\omega_{L_{1}}^{2}[w_{1}(r\cos{\theta})]^{2}}\bigg[\frac{2r^{2}\sin^{2}{\theta}}{w_{1}(r\cos{\theta})^{2}}\bigg]^{|m|}\\
\times\exp\bigg[\frac{-2r^{2}\sin^{2}{\theta}}{w_{1}(r\cos{\theta})^{2}}\bigg]+\frac{e^{2}I_{2}w^{2}_{0_{2}}}{2|m|!m_{e} c\epsilon_{0}\omega_{L_{2}}^{2}[w_{2}(r\cos{\theta})]^{2}}\\\times\bigg[\frac{2r^{2}\sin^{2}{\theta}}{w_{2}(r\cos{\theta})^{2}}\bigg]^{|m|}\exp\bigg[\frac{-2r^{2}\sin^{2}{\theta}}{w_{2}(r\cos{\theta})^{2}}\bigg].
\end{multline}
 The trapping potential $V_{p}$, visualized in Fig.~\ref{fig:potentials}(b), acts on the quasi-free Rydberg electron and radially traps the atoms within the center of the modes, for the intensity scales as $\rho^{2m}$ as $\rho\rightarrow 0$. In Fig.~\ref{fig:trapping_potential}, we plot a cut through $V_{p}$ for laser modes with wavelengths $\lambda_{1}=536$ nm and $\lambda_{2}=532$ nm.

When making the rotating-wave approximation, we neglected several terms in the $A^{2}$ part of the Hamiltonian that are highly energy-non-conserving on optical energy scales; these terms correspond the absorption of photon pairs or emission of photon pairs. 
\subsection{Rabi frequency}
An S-state Rydberg atom's radius scales as $2n^{2}a_{0}$. However, as the angular momentum of a Rydberg atom increases, its radius decreases. For a CS, the radius is $n^{2}a_{0}$. Therefore, it is not feasible to circularize Rydberg atoms of the same $n$ in a single step due to small wave function overlap. For optimal Rabi frequencies, one must choose a CS of a principal quantum number $n'$ and a low-$m_{l}$ Rydberg atom of principal quantum number $n\simeq n'/\sqrt{2}$. For this calculation, we chose to transfer $\ket{g}=\ket{21\text{F}_{7/2},m_{l}=3}$ atoms to $\ket{e}=\ket{n=32,l=31,m_{l}=31}$. The overlap of the radial wave functions $U_{g}(r_{e}),U_{e}(r_{e})$ of these two states is exhibited in Fig. 3.
\begin{figure}[h!]
\begin{center}
\includegraphics[scale=0.7]{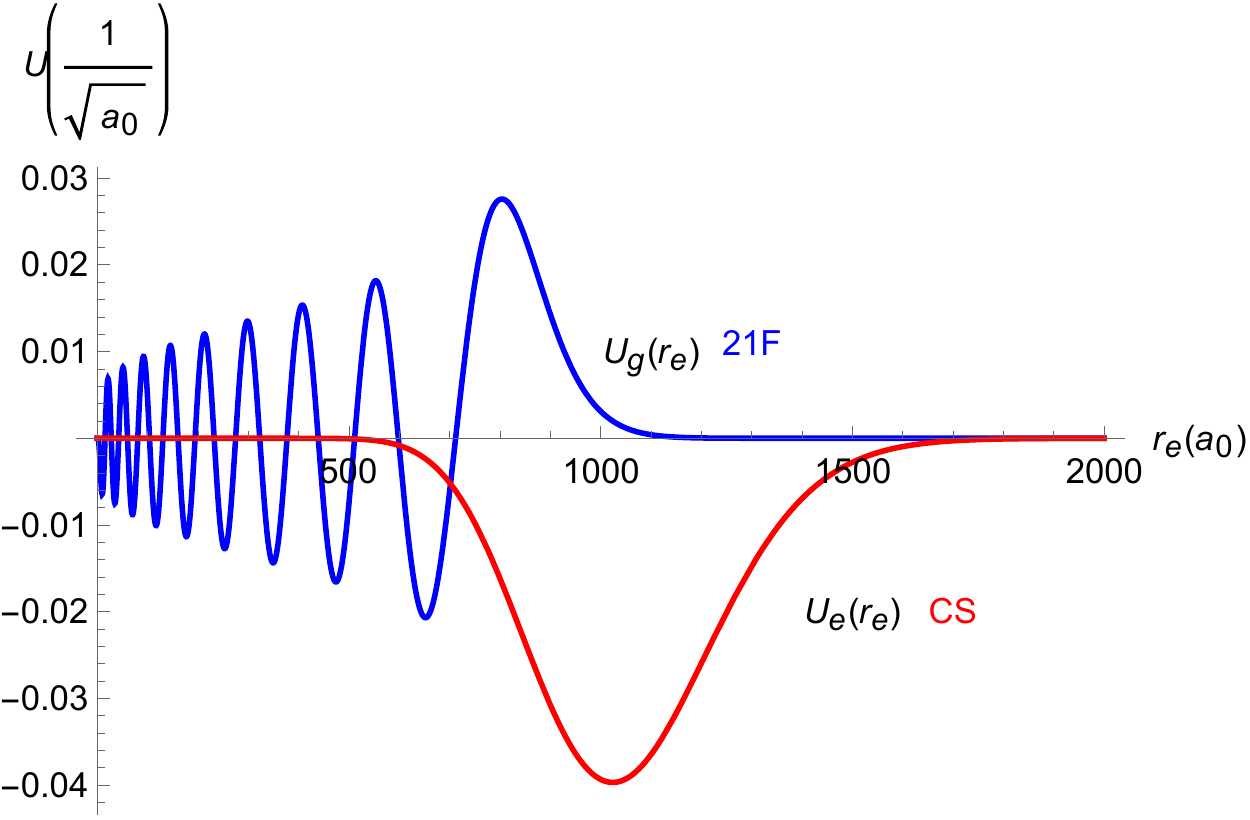}\\

\caption{Radial wave function overlap of the states discussed in the text: $\ket{g}$ (blue) and $\ket{e}$ (red).}
\label{fig:wave_function_overlap}
\end{center}
\end{figure}

The energy splitting between $\ket{g}$ and $\ket{e}$ is $h\times$4.2~THz. That means we must choose LG modes with frequencies such that $\omega_{b}=2\pi\times4.2$ THz. We choose to model an experiment in which $\lambda_{1}$, the wavelength of LG mode 1 is 536 nm, and $\lambda_{2}=532$~nm, which are both far off-resonant from any transition in $^{85}$Rb, the alkali we use. These co-propagating LG modes give rise to a ponderomotive interaction term $V_{C}(\textbf{r},t)$ in the Hamiltonian (see Eq.~\ref{eq:circ}).
In order to calculate the Rabi frequency for this transition as a function of atomic center-of-mass position $\textbf{R}$, we must obtain matrix element $\bra{e}V_{C}(\textbf{R}+\textbf{r}_{e})\ket{g}$ by integrating over the relative Rydberg-electron coordinate $\textbf{r}_{e}$. Thus, we calculate the Rabi frequency
\begin{equation}
\Omega(\textbf{R})=\frac{2}{\hbar}\int \psi_{e}^{*}(\textbf{r}_{e})V_{C}(\textbf{R}+\textbf{r}_{e})\psi_{g}(\textbf{r}_{e})d^{3}r_{e},
\end{equation}
where $\psi_{g}(\textbf{r}_{e})$ and $\psi_{e}(\textbf{r}_{e})$ are the electronic wave functions of $\ket{g}$ and $\ket{e}$, respectively. Due to negligible quantum defects for both states, the radial wave functions can be assumed to be hydrogenic. For our calculations, we use the parameters of optical power $P_{0_{1}}=P_{0_{2}}=150$ mW and waists $w_{0_{1}}=3.41\text{ }\mu$m and $w_{0_{2}}=3.39\text{ }\mu$m. The Rabi frequencies presented in Fig.~\ref{fig:Rabi_adiabatic} are numerically integrated with a quasi-Monte Carlo algorithm.

\subsection{Adiabatic potentials}
\par In general, states $\ket{g}$ and $\ket{e}$ will see different ponderomotive energy shifts introduced by $V_{p}(\textbf{R}+\textbf{r}_{e})$. These energy shifts are on the order of the Rabi frequency and are responsible for a cylindrical trap for the Rydberg atoms. Through the application of external magnetic and electric fields, it can be ensured that the perturbations of the ponderomotive potential are much less than the external-field-induced frequency splittings among the relevant atomic states. Thus, we can assume that the electron's wave function is the same for all center-of-mass positions $\textbf{R}$. In order to determine the ponderomotive energy shift as a function of center-of-mass position for the CS, we calculate the Born-Oppenheimer adiabatic potential from non-degenerate perturbation theory,
\begin{equation}
\label{eq:adiabatic}
    V_{ad}(\textbf{R})=\int \psi^{*}_{e}(\textbf{r}_{e})V_{p}(\textbf{R}+\textbf{r}_{e})\psi_{e}(\textbf{r}_{
    e})d^{3}r_{e}.
\end{equation} 

\begin{figure}[h!]
\begin{center}
\includegraphics[scale=0.65]{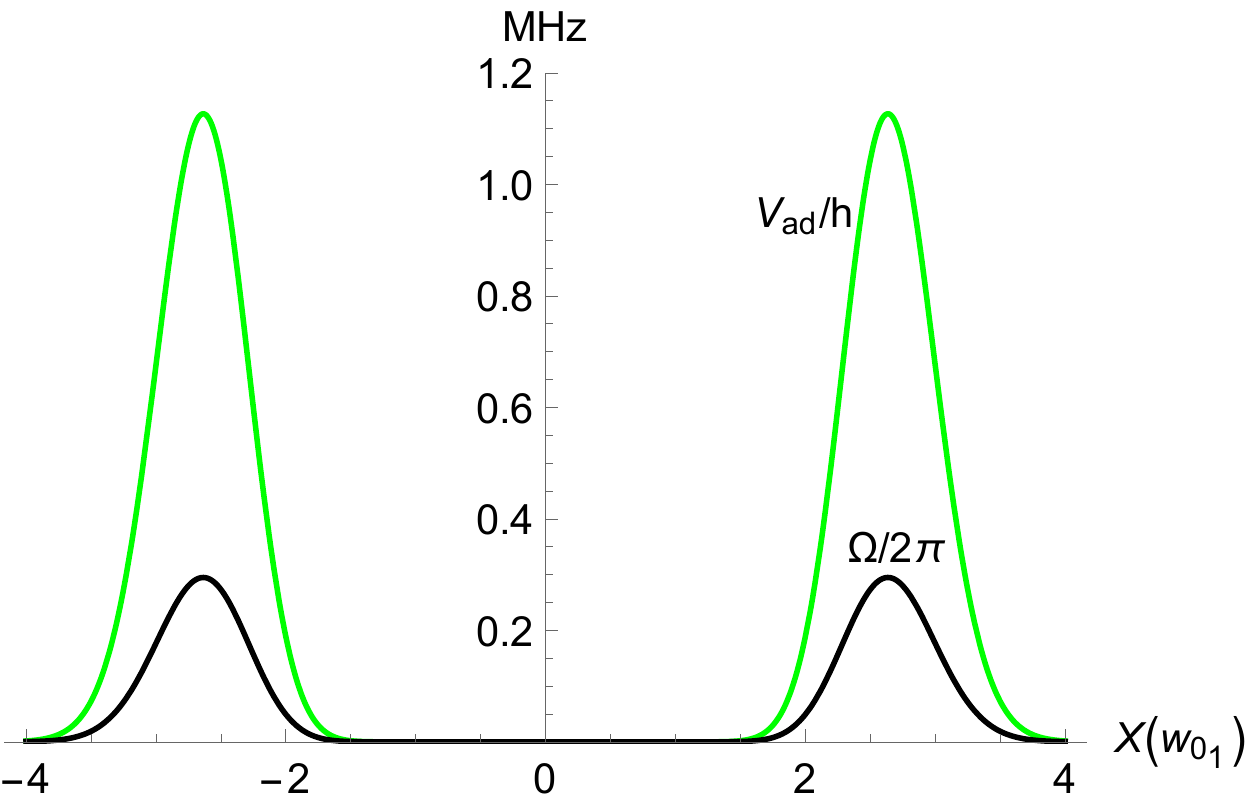}
\caption{Adiabatic potential (green) seen by the CS atom of $n=32$ as a function of the atom's center-of-mass position along the $X$-axis for $w_{0_{1}}=3.41\text{ }\mu$m $w_{0_{2}}=3.39\text{ }\mu$m and $P_{0_{1}}=P_{0_{2}}=150$ mW. The Rabi frequency (black) for the transition of 21F to the CS of $n=32$ along the $X$-axis for the same optical parameters.}
\label{fig:Rabi_adiabatic}
\end{center}
\end{figure}

Numerical integration of Eq.~\ref{eq:adiabatic} with a quasi-Monte Carlo algorithm results in the energy shifts shown in Fig.~\ref{fig:Rabi_adiabatic} for $\ket{e}$ along the $X$-axis. Because the diameters of the atoms in states $\ket{g}$ and $\ket{e}$ are $\sim90$~nm and the diameter of the radial trap is $\sim20\text{ }\mu$m, these low-$n$ Rydberg atoms are effectively point-like particles with respect to the trap's intensity profile. Therefore, through calculation of Eq.~\ref{eq:adiabatic}, there is little spatial averaging over the electron's wave function and $V_{ad}\simeq V_{p}$ in magnitude. As a result, $\ket{g}$ and $\ket{e}$ see the same adiabatic potential.
\subsection{Experimental considerations}
\par The difference in level shifts $\Delta V_{ad}$, i.e., the adiabatic potential of $\ket{g}$ subtracted from that of $\ket{e}$, has a dependence on $\textbf{R}(t)$, the center-of-mass position as a function of time. However, as mentioned above, because the wave functions of $\ket{g}$ and $\ket{e}$ have similar sizes, the trap is nearly magic, as it has a maximum $\Delta V_{ad}$ of $0.7$ kHz. In order to evaluate the circularization efficiency, it is required to consider the trajectories of the trapped atoms. We can extract $\textbf{R}(t)$ from Newton's equations by treating the atoms as classical bodies in a two-dimensional trapping potential. The trapped atoms can be modeled by an ensemble with a uniform spatial distribution and a Maxwell-Boltzmann distribution of speed. A typical rms-speed of an atom in a Rb corkscrew molasses is 5~cm/s. Simulations show that this case would lead to a typical round-trip period of  0.6 ms for the aforementioned parameters. This period is almost 60 times the lifetime of an atom in the 21F state in a 300 K blackbody field, meaning that the atoms excited into the 21F state with an initial center-of-mass position at the center of the trap would decay before they reach the trap walls (where they would be circularized if they were still in 21F). Therefore, only the atoms initially close to the edge of the trap have a chance of becoming circularized. We estimate the total fraction of the ensemble that is circularized to be about 5\%. 

The energy splitting between $\ket{g}$ and $\ket{e}$ is several THz in order to guarantee sufficient wave function overlap.  
It is not trivial to phase-lock two lasers with a THz frequency difference. A possible realization of this scheme is the following. Two tunable lasers at $532$ nm (beam 1) and $536$ nm (beam 2) are phase-locked to two modes of a frequency comb laser separated by $4.2$ THz and diffract off spatial light modulators (SLMs) or digital micro-mirror devices (DMDs) to give opposite winding numbers $14$ and $-14$~\cite{Mirhosseini2013}. When the beams are combined, they overlap a cold sample of $^{85}$Rb, prepared in an optical molasses at $\sim10\text{ }\mu$K. Cold atoms are adiabatically loaded into the radial trap by slowly ramping up beam 1 to allow efficient cooling within the center of the LG beams, while keeping beam 2 off. The atoms are then optically pumped into the $\ket{5\text{S}_{1/2},m_{F}=3}$ Zeeman sublevel with a $780$ nm laser of $\sigma^{+}$-polarization. Subsequently, the sample is excited to $\ket{g}$ with 780 nm, 776 nm, and 1292 nm lasers of $\sigma^{+}$-polarizations. Beam 2 is then pulsed on for a duration that allows optimal transfer into the CS. The timing and pulse shapes of this procedure may be optimized by quantum optimal control theory to yield the highest fidelity~\cite{Signoles2017, Patsch2018}.

\section{Rapid adiabatic passage in an RF-modulated POL}
In the previous section, we considered a case where an atom is circularized by a single ponderomotive interaction that is highly forbidden for an electric-dipole interaction ($\Delta m_{l}= 28$). For this section, we consider many quadrupole-like ponderomotive interactions that lead to circularization in a RAP scheme~\cite{Nussenzveig193,Lutwak1997}. We consider an optical lattice that is a superposition of a laser beam shifted in frequency by $\omega_{rf}$ and an unmodulated beam. We can use the effects of the ponderomotive interaction~\cite{Knuffman2007, Moore2015} to couple states for the RAP method. For this calculation, we consider the hydrogenic states of the $n=32$ Rydberg level under static, homogeneous fields $F=2.736$ V/cm and $B=5.0$ G. Also, we assume the carrier frequency of the laser is $532$ nm. The bound-state energy $W$, in the parabolic basis of $\ket{n,n_{1},n_{2},m_{l}}$, with $n=n_{1}+n_{2}+|m_{l}|+1$, is given by
\begin{multline}
    W=-hc\bigg(\frac{m_{+}}{m_{e}+m_{+}}\bigg)\frac{R_{\infty}}{n^{2}}\\+\frac{3}{2}Fea_{0}n(n_{1}-n_{2})+\frac{e\hbar B}{2m_{e}}(m_{l}+g_{s}m_{s})\\-\frac{1}{4}\pi \epsilon_{0} a_{0}^{3} F^{2} n^{4}[17n^{2}-3(n_{1}-n_{2})^{2}-9m_{l}^{2}+19],
\end{multline}
where $m_{+}$ is the mass of the $^{85}$Rb ion core, $R_{\infty}$ is the Rydberg constant, $g_{s}=2$, and $m_{s}$ is the Zeeman sublevel of the electron's spin. The last term of $W$ is responsible for modeling the energy splittings in the quadratic Stark effect.
\par While the static fields are polarized along the quantization-axis $\hat{z}$, the POL propagates along $\hat{x}$ and is polarized along $\hat{y}$. It is important to note that because the theory of RAP is best-described in the dressed-atom picture, we represent the POL laser's vector potential as
\begin{equation}
\textbf{A}(\textbf{r},t)=\hat{y}\sum_{\textbf{k}}\frac{\mathcal{E}_{\textbf{k}}}{2\omega_{\textbf{k}}}a_{\textbf{k}}e^{i(\textbf{k}\cdot\textbf{r}-\omega_{\textbf{k}}t)}+\text{h.c.},
\end{equation}
with $\mathcal{E}_{\textbf{k}}$ being the quantized field amplitude $\bigg(\sqrt{\frac{2\hbar \omega_{\textbf{k}}}{\epsilon_{0}\mathcal{V}}}\bigg)$ and $a_{\textbf{k}}$ being the annihilation operator for mode $\textbf{k}$. Recall that $\textbf{r}$ is the position vector in the laboratory frame that is the sum of the atom's center-of-mass position $\textbf{R}$ and the Rydberg electron's relative coordinate $\textbf{r}_{e}$. For the case of a modulated POL, $\textbf{k}=\pm\textbf{k}_{1},\pm\textbf{k}_{2}$
with $\textbf{k}_{1}$ being the wave vector corresponding to the unmodulated mode with angular frequency $\omega_{L}$ and $\textbf{k}_{2}$ corresponding to the mode modulated by $\omega_{rf}(t)$. The minus signs correspond to the backwards-propagating modes. The angular frequency accompanying mode $\textbf{k}$ is denoted by $\omega_{\textbf{k}}$. If we consider the atom in the presence of parallel electric and magnetic fields, $F$ and $B$, respectively, we can make the Born-Oppenheimer approximation by adiabatically separating $\textbf{R}$, the center-of-mass position in the laboratory frame, from $\textbf{r}_{e}$, the electron's position in the atom's frame. Under the assumptions of real field amplitudes and a perfectly balanced lattice, the quasi-free electron ponderomotive term $\frac{e^{2}A(\textbf{r}=\textbf{R}+\textbf{r}_{e},t)^{2}}{2m_{e}}$ gives an interaction potential
$V_{AF}(\textbf{R}+\textbf{r}_{e},t)$ described by
\begin{multline}
\label{eq:atom_field}
    V_{AF}(\textbf{R}+\textbf{r}_{e},t)=\frac{e^{2}f(t)\mathcal{E}_{\textbf{k}_{2}}\mathcal{E}_{\textbf{k}_{1}}}{4m_{e}\omega_{L}[\omega_{L}+\omega_{rf}(t)]}\\
    \times\{a_{\textbf{k}_{2}}a^{\dag}_{-\textbf{k}_{1}}
   \exp[i(k_{1}+k_{2})(X+x_{e})]\\
  +a_{-\textbf{k}_{2}}a^{\dag}_{\textbf{k}_{1}}\exp[-i(k_{1}+k_{2})(X+x_{e})]\}
\end{multline}
where $a_{\pm\textbf{k}_{2}}$ and $a^{\dag}_{\pm\textbf{k}_{1}}$ are, respectively, the annihilation and creation operators for a modulated mode and an unmodulated mode. The modulated mode also has a temporal envelope described by $f(t)$. Because the rf frequency is so much smaller than the optical frequency of the lattice, we can neglect the $\omega_{rf}(t)$ in the denominator of Eq.~\ref{eq:atom_field}. Note that we adopt the rotating-wave approximation to arrive at Eq.~\ref{eq:atom_field}.
\par
\begin{figure*}[h!]
\begin{center}
\includegraphics[scale=0.50]{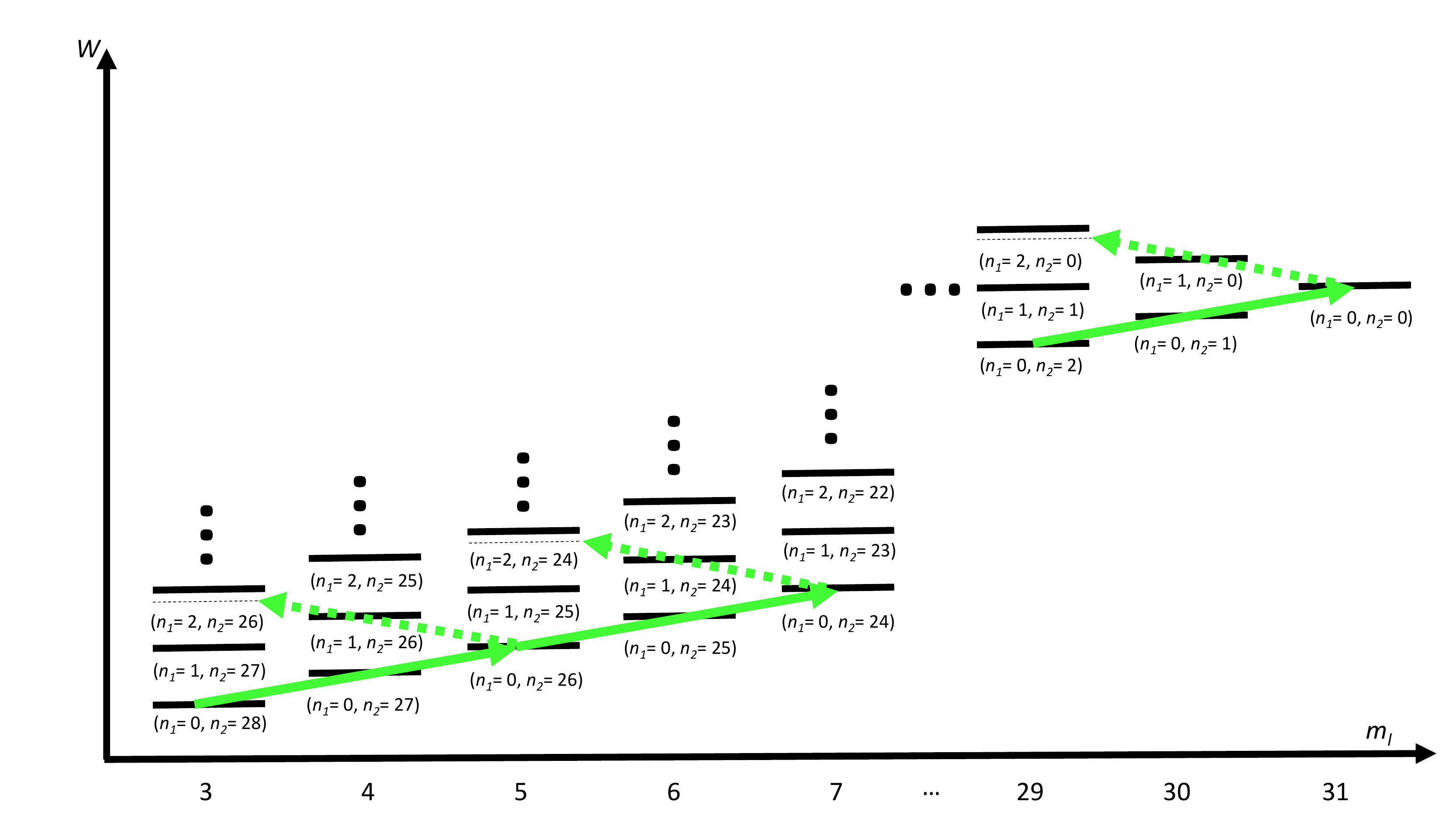}

\caption{Hydrogenic manifold in the parabolic basis $n=n_{1}+n_{2}+|m_{l}|+1$ under parallel electric and magnetic fields $F$ and $B$. The solid green arrows represent desired couplings provided by the rf-modulated POL; the dashed lines represent undesired ``leakage" transitions that  reduce the overall circularization efficiency. Magnetic field $B$ detunes these ``leakage" transitions from resonance and minimizes CS population loss from them. }
\label{fig:levels}
\end{center}
\end{figure*}

 The interaction modeled by Eq.~\ref{eq:atom_field} describes the inelastic scattering process of a forward-propagating modulated photon into an unoccupied backwards-propagating mode. During the scattering process, the modulated photon imparts its rf energy to the atom, promoting it to a state with $\Delta m_{l}=2$, as shown in Fig.~\ref{fig:levels}.
\par The POL interacts with hydrogenic states of a principal quantum number $n$ with Zeeman sublevels $m_{l}$ and parabolic numbers $n_{1}=0$, and $n_{2}=n-1-m_{l}$. Additionally, because we work in the dressed-atom picture involving photons of modes $\pm\textbf{k}_{2}$ and $\mp\textbf{k}_{1}$, we must include their photon numbers, which we will represent by $N+n_{2}/2$ and $M-n_{2}/2$, respectively where $N$ and $M$ are background photon numbers that can be set to zero in the energy eigenvalues, as they contribute the same offset for each state. Note that if we assume perfectly-balanced lattice, there are the same number of photons for modes $\textbf{k}$ and $-\textbf{k}$.  Thus, we characterize the dressed states with $\ket{i}=\ket{m_{l},n_{2}=n-1-m_{l},N+n_{2}/2,M-n_{2}/2}$ and eigenvalues
\begin{equation}
    W'_{i}(t)=W+\frac{n_{2}}{2}\hbar(\omega_{rf}+\alpha t)+V_{ad,n_{2},m_{l}},
\end{equation}
where $V_{ad,n_{2},m_{l}}$ is the offset of the ponderomotive lattice  shift, which is determined to differ among states within $n=32$. The modulator imprints a phase of $-\frac{1}{2}(2\omega_{rf,0} t +\alpha t^{2})$ on the transmitted wave for a POL that is chirped with a frequency range  $\Delta\nu$ over a period $\tau$, where $\alpha=2\pi\frac{\Delta\nu}{\tau}$~\cite{Malinovsky2001} and $\omega_{rf,0}$ is the rf center frequency.
\par The relevant eigenstates for $n=32$ and $n_{1}=0$ are represented as coherent superpositions of spherical hydrogenic states given by
\begin{multline}
    \ket{m_{l},n_{2},N+n_{2}/2,M-n_{2}/2}=\\
    \sum_{l}(-1)^{(-31+m_{l}-n_{2})/2+l}\sqrt{2l+1}\\
    \times\bigg(\begin{matrix}
     \frac{31}{2} & \frac{31}{2} & l  \\
     \frac{m_{l}-n_{2}}{2} & \frac{m_{l}+n_{2}}{2} & -m_{l}
\end{matrix}\bigg)\ket{n=32,l,m_{l}}\\
\otimes\ket{N+n_{2}/2,M-n_{2}/2},
\end{multline}
where the terms in parentheses are the Wigner 3-j symbol~ \cite{Gallagher}.
\par In the limit of strong laser fields, the interaction potential becomes
\begin{multline}
    V_{AF}(\textbf{R}+\textbf{r}_{e},t)=\frac{e^{2}e^{-\ln{(16)}t^{2}/2\tau^{2}}\sqrt{I_{\textbf{k}_{1}}I_{\textbf{k}_{2}}}}{m_{e}c\epsilon_{0}\omega_{L}^{2}}\\
    \times\cos{[(k_{1}+k_{2})(X+x_{e})]}
\end{multline}
Also, we set $f(t)=e^{-\ln{(16)}t^{2}/2\tau^{2}}$.
An atom sitting at the bottom of a potential well will see a quadratic potential approximated by
\begin{multline}
    V_{AF}(\textbf{r}_{e})\simeq\frac{e^{2}e^{-\ln{(16)}t^{2}/2\tau^{2}}\sqrt{I_{\textbf{k}_{1}}I_{\textbf{k}_{2}}}(k_{1}+k_{2})^{2}}{2m_{e}c\epsilon_{0}\omega^{2}_{L}}r_{e}^{2}\\
\times    \bigg[\sqrt{\frac{2\pi}{15}}Y^{2}_{2}(\theta_{e},\phi_{e})
    +\sqrt{\frac{2\pi}{15}}Y^{-2}_{2}(\theta_{e},\phi_{e})+\frac{\sin^{2}{\theta_{e}}}{2}\bigg],
\end{multline}
which corresponds to a non-diagonal matrix element of
\begin{multline}
\bra{m_{l}+2,n_{2}-2}V_{AF}(\textbf{r}_{e})\ket{m_{l},n_{2}}\simeq\frac{e^{2}e^{-\ln{(16)}t^{2}/2\tau^{2}}}{4m_{e}c\epsilon_{0}\omega^{2}_{L}}\\
\times\sqrt{2I_{\textbf{k}_{1}}I_{\textbf{k}_{2}}/3}(k_{1}+k_{2})^{2}\sum_{l,l'}(-1)^{-31-n_{2}+l+l'}(r_{e}^{2})^{l'}_{l}\\
\times(2l+1)(2l'+1)\bigg(\begin{matrix}
     \frac{31}{2} & \frac{31}{2} & l'  \\
     \frac{m_{l}-n_{2}}{2}+2 & \frac{m_{l}+n_{2}}{2} & -m_{l}-2
\end{matrix}\bigg)\\
\times\bigg(\begin{matrix}
     l & 2 & l' \\
     m_{l} & 2 & -m_{l}-2
\end{matrix}\bigg)\bigg(\begin{matrix}
     l & 2 & l' \\
     0 & 0 & 0
\end{matrix}\bigg)\bigg(\begin{matrix}\frac{31}{2} & \frac{31}{2} & l  \\
     \frac{m_{l}-n_{2}}{2} & \frac{m_{l}+n_{2}}{2} & -m_{l}\end{matrix}\bigg),
\end{multline}
where $(r_{e}^{2})^{l'}_{l}$ is the radial matrix element between two hydrogenic states in the spherical basis.
\par  Non-adiabatic transitions that reduce the efficiency of RAP schemes are best modeled using the Schr{\"o}dinger equation in the adiabatic basis. The adiabatic eigenkets $\ket{j}$ of this Hamiltonian can be obtained by applying a unitary transformation~\cite{Rubbmark1981,Berman} $D$, where
\begin{equation}
    \ket{j}=\sum_{i}D_{ij}\ket{i}.
\end{equation}
Additionally,
\begin{equation}
    \ket{i}=\sum_{j}D^{*}_{ji}\ket{j}.
\end{equation}
Thus, for quantum state $\ket{i}$ with Schr{\"o}dinger equation
\small
\begin{align}
    \sum_{j}\bigg[i\hbar\partial_{t}(D^{*}_{ji}\ket{j})&=HD^{*}_{ji}\ket{j}\bigg]\\
    \sum_{j}\bigg[i\hbar\dot{D}^{*}_{ji}\ket{j}+i\hbar D^{*}_{ji}\partial_{t}\ket{j}&=\hat{H}D^{*}_{ji}\ket{j}\bigg],
\end{align}
\begin{multline}
     \sum_{j}\bigg[\bra{j'}i\hbar D_{ij'}\dot{D}^{*}_{ji}\ket{j}+\bra{j'}i\hbar D_{ij'}D^{*}_{ji}\partial_{t}\ket{j}\\=\bra{j'}D_{ij'}HD_{ji}^{*}\delta_{jj'}\ket{j}\bigg],
\end{multline}
\normalsize
where the $\bra{j'}i\hbar D_{ij'}\dot{D}^{*}_{ji}\ket{j}$ term is responsible for non-adiabatic transitions from one adiabatic ket, $\ket{j}$, to another, $\ket{j'}$. The idea behind the RAP method is to minimize this term such that a state initialized at $m_{l}=3$ arrives at $m_{l}=31$ at the end of the frequency chirp. Efficiency of this process is diminished if atoms are lost to other adiabatic states via non-adiabatic transitions.

\begin{figure}[h!]
\begin{center}
\includegraphics[scale=0.6]{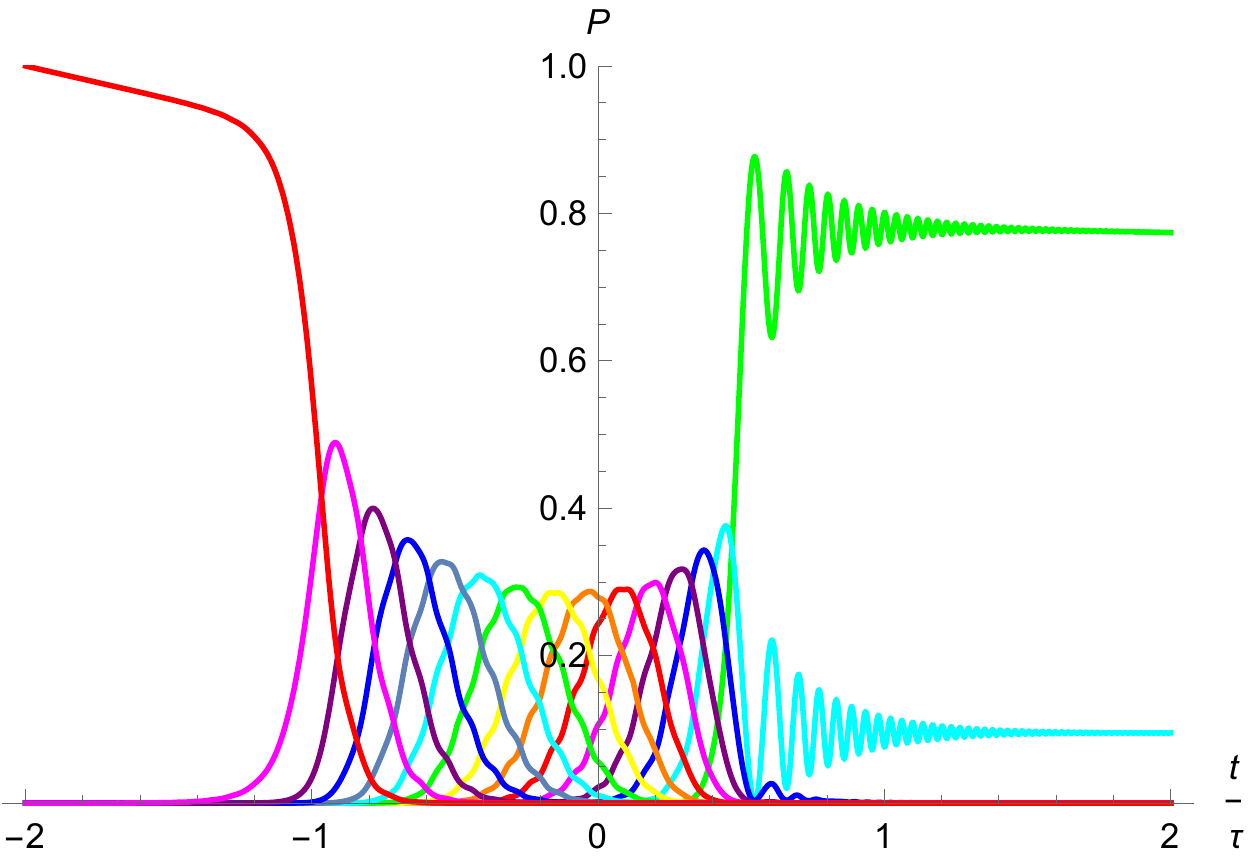}\\
\includegraphics[scale=0.6]{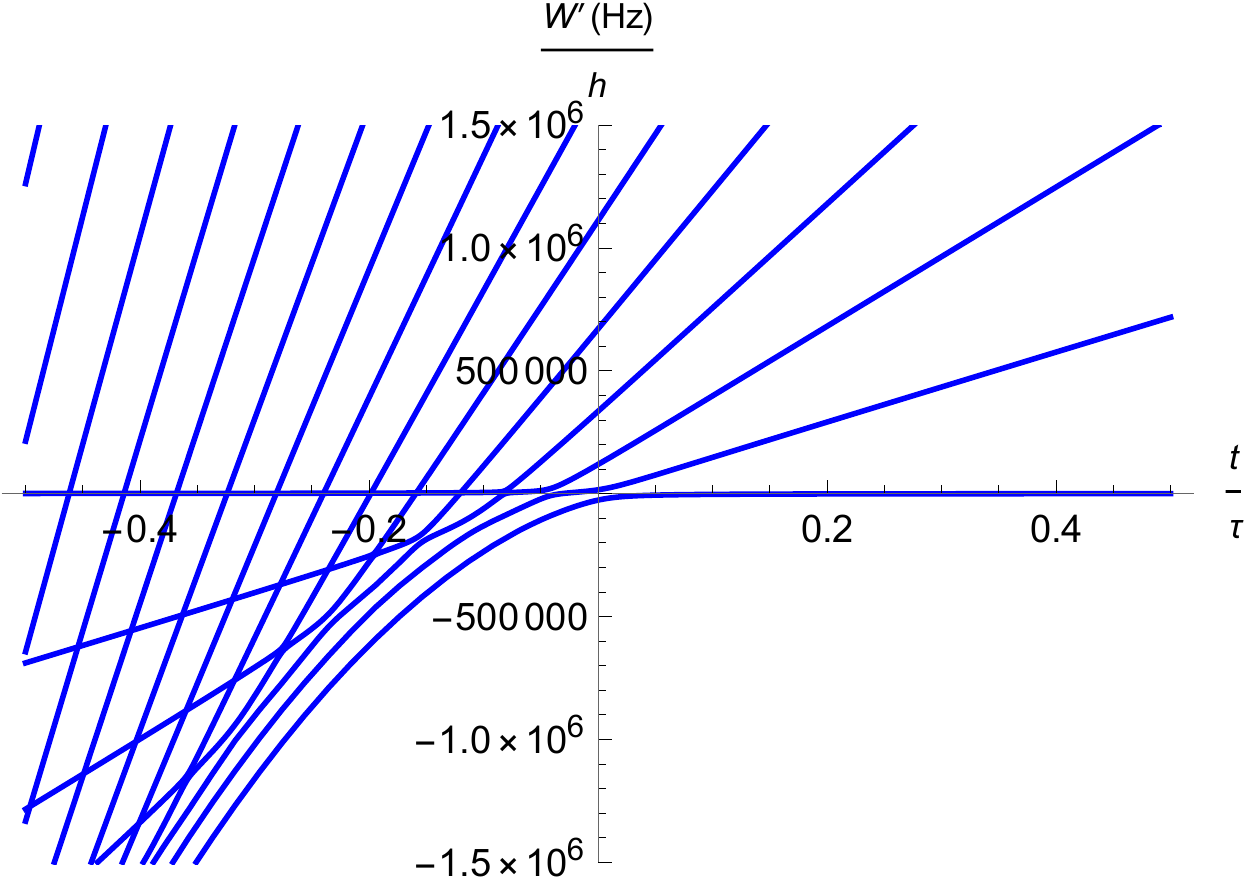}
\caption{The top figure displays the probability of atoms populating hydrogenic states from $\ket{3,28,N+14,M-14}$ to the CS, $\ket{31,0,N,M}$, as a function of time scaled by $\tau=25\text{ }\mu$s and $\Delta\nu=1.41$ MHz. The right-most green curve displays the probability of atoms populating the CS, while the left-most red curve represents the population of atoms in the $\ket{3,28,N+14,M-14}$ state. The bottom figure shows the adiabatic eigenenergies of the Hamiltonian dressed by the inelastic scattering interactions of the lattice photons.}
\label{fig:adiabatic}
\end{center}
\end{figure}

In the regime in which the Rabi-frequency is lower than the splittings of the second-order Stark effects and ponderomotive shifts, the RAP involves sequential two-level Landau-Zener transitions into the CS. The probability of atoms in state $\ket{j}$ transitioning to $\ket{j'}$ is, as presented in~\cite{Rubbmark1981,Zener1932},
\begin{equation}
\label{eq:diabatic_rate}
    P(j\rightarrow j')=e^{-2\pi\Gamma},
\end{equation}
where
\begin{equation}
    \Gamma=\frac{|\bra{i'}V_{AF}\ket{i}|^{2}}{\hbar |d[W_{i'}'(t)-W_{i}'(t)]/dt|}.
\end{equation}
In order to achieve a non-adiabatic transition probability lower than $\sim0.01$, $2\pi\Gamma>>4$. However, we calculate Rabi frequencies as high as $2\pi\times1.90$ MHz for the case of a perfectly-balanced lattice consisting of 1.43 W modulated and unmodulated beams focused to a waist of 10$\text{ }\mu m$. These coupling strengths put us in a regime where the atom transitions into multiple states as a time, as shown in Fig.~\ref{fig:adiabatic}, making two-level Landau-Zener models inaccurate.
\par Fig.~\ref{fig:adiabatic} shows the population $P$ of atoms occupying the lowest ladder of the hydrogenic manifold in parabolic coordinates for $\tau=25\text{ }\mu$s and $\Delta\nu=1.41$ MHz for a Gaussian amplitude modulation and a linear chirp. Notice that the beginning of the chirp rate seems to have the behavior of a two-level Landau-Zener transition. That is because there exists a differential ponderomotive shift $\Delta V_{ad}$ between atoms in $m_{l}=3$ and those in $m_{l}=5$ is comparable to the Rabi frequency of the coupling between them, i.e., $\Omega=2\pi\times 1.04$ MHz and $\Delta V_{ad}/\hbar = 2\pi\times 1.28$ MHz for atoms situated at the bottom of a lattice well. 
Throughout the middle of the RAP procedure, the 
Rabi frequencies are much larger and allow multiple non-adiabatic states to be excited at the same time. We find a circularization efficiency from this simulation to be $89\%$.
\par While $n=32$ Rydberg atoms with low angular momenta would decay from radiative losses after $\sim50\text{ }\mu$s in a $4$ K environment, CS atoms shielded from thermal photons would live much longer, at the order of $\sim10$ ms. The development of a nonlinear chirp that maintains an average rate equal to $\alpha$, yet scans more quickly during the passage through low-$m_{l}$ states than the rate through high-$m_{l}$ states, would circumvent this issue.
\par The ``leakage" transitions represented by the dashed lines in Fig.~\ref{fig:levels} lead to a significant reduction of the atoms in the CS at the end of the sequence. Typical RAP schemes for circularization require the application of a magnetic field to lift the degeneracy of the unwanted and desired transitions~\cite{Nussenzveig193}. For our calculation, where these transitions differ in resonant frequency by $\sim28$ MHz, we do not expand our Hilbert space to account for such couplings.
\par The experimental realization of this method would require constructing a POL with a high-powered laser beam split in a Mach-Zehnder interferometer with one arm acousto-optically modulated by an rf source of center angular frequency $\omega_{rf,0}=2\pi\times350$ MHz and the other unmodified. Suitable electrode and Helmholtz coil geometries would allow the static fields used in this calculation.
\section{Circularization of Rydberg atoms in a time-orbiting ponderomotive optical lattice}
\par As previously mentioned, the fidelity of circularization schemes is reduced by unwanted transitions within the hydrogenic manifold if the polarization of the coupling is not purely $\sigma^{+}$ or $\sigma^{-}$. In this section, we will show that a time-orbiting ponderomotive optical lattice (TOPOL), with effective electro-optic control, would provide potentials that would drive transitions equivalent to those of purely $\sigma^{\pm}$-polarized rf radiation.
\par The idea of the TOPOL is that the two-dimensional optical lattice is constructed by a ponderomotive potential with a rapidly-orbiting, time-dependent component, resulting in a static-part that is approximately harmonic and a time-dependent part that is equivalent to rotating rf electric field. Such a potential is realized by a two-dimensional POL with the $x$ and $y$-components phase shifted with a cosine and sine-like time dependence, respectively.  In order to drive transitions between two Rydberg states, the phase-modulation frequency must be equal to the resonant frequency of a Rydberg transition. It will be shown that the effect is an electric-dipole coupling between Rydberg states in a manner such that $m_{l}$ can only increase or decrease by one unit but not both.
\par Consider the intersection of four optical fields described by
\begin{align}
\textbf{E}^{(+)}_{1}(x,t)&=\hat{\epsilon}^{(1)}\mathcal{E}^{(+)}_{1}\cos{\big[kx-\omega_{L}t+\beta_{x}\cos{(\omega_{\text{rf}}t)}\big]},\\
\textbf{E}^{(+)}_{2}(y,t)&=\hat{\epsilon}^{(2)}\mathcal{E}^{(+)}_{2}\cos{\big[ky-\omega_{L}t+\beta_{y}\sin{(\omega_{\text{rf}}t)}\big]},\\
\textbf{E}^{(-)}_{1}(x,t)&=\hat{\epsilon}^{(1)}\mathcal{E}^{(-)}_{1}\cos{\big(kx+\omega_{L}t\big)},\\
\textbf{E}^{(-)}_{2}(y,t)&=\hat{\epsilon}^{(2)}\mathcal{E}^{(-)}_{2}\cos{\big(ky+\omega_{L}t\big)},
\end{align}
where $x=X+x_{e}$ and $y=Y+y_{e}$ are coordinates in the laboratory frame and $\beta_{x}$ ($\beta_{y}$) is the amplitude of the phase shift of the beam forward-propagating along $\hat{x}$ ($\hat{y}$). Here, we assume that $\beta_{x}=\beta_{y}=\beta$. We use the assumptions that the polarization vectors $\hat{\epsilon}^{(1)}$ and $\hat{\epsilon}^{(2)}$ are orthogonal and that the counterpropagating beams have polarizations parallel to those of the forward-propagating beams. Time integration of these fields yields the vector potential operators used in the minimal coupling Hamiltonian.
\par
\begin{figure}[h!]
\begin{center}
\includegraphics[scale=0.55]{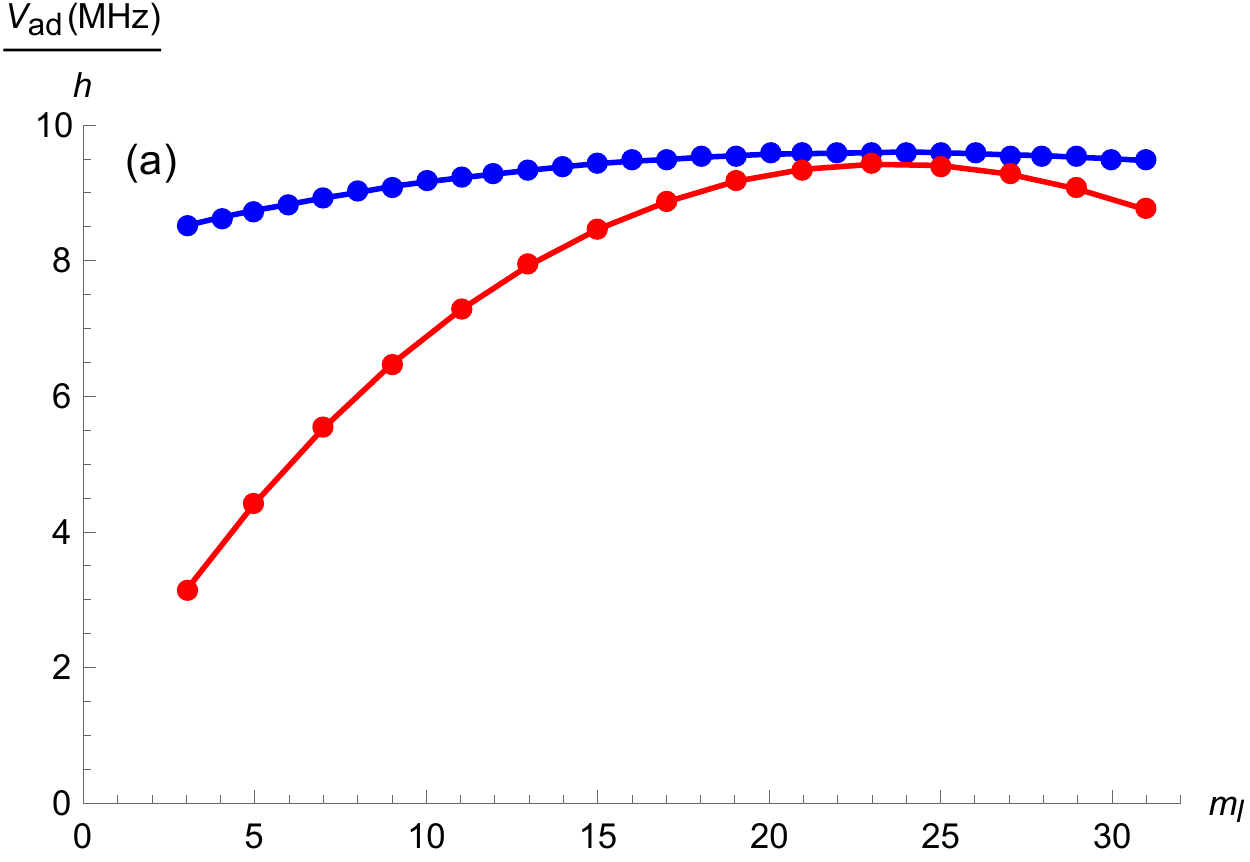}\\
\includegraphics[scale=0.55]{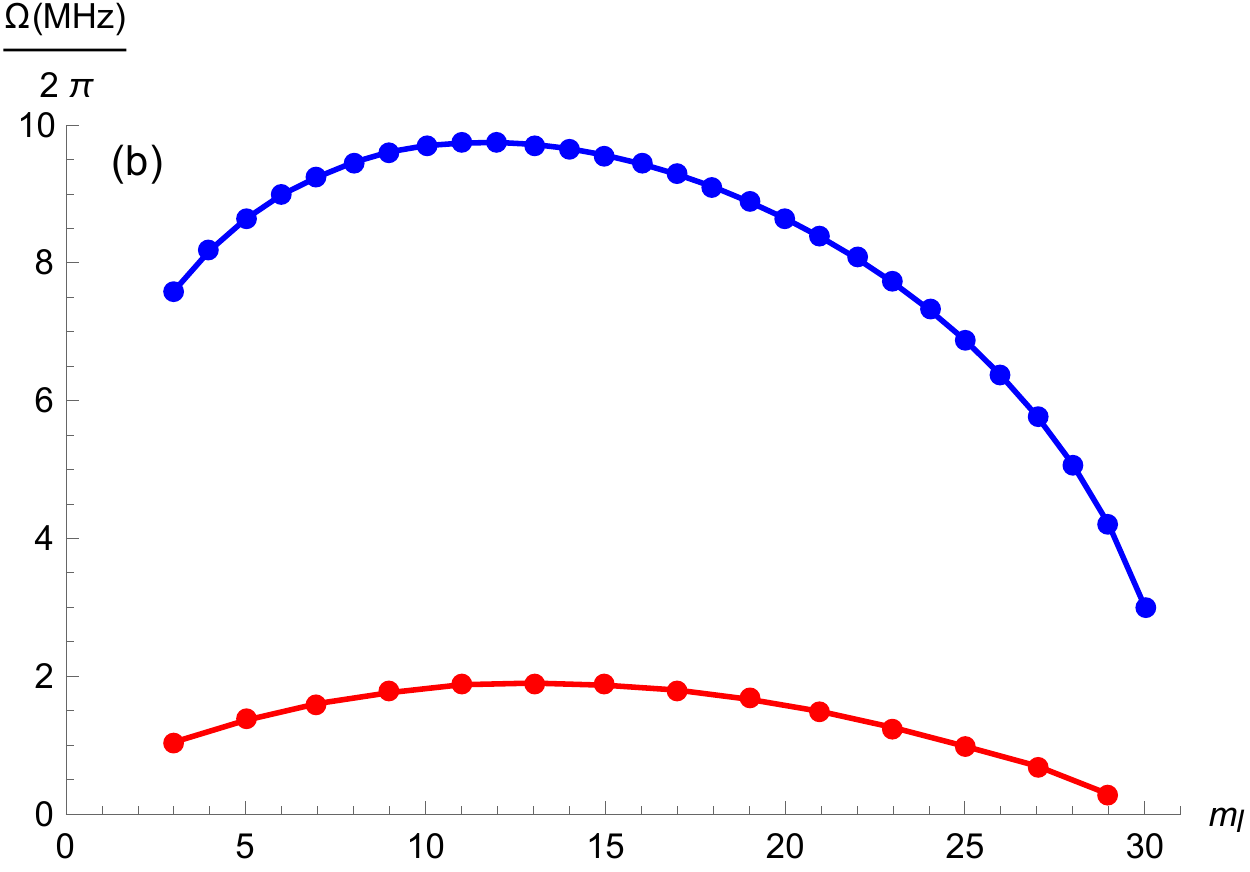}
\caption{ (a) Ponderomotive shifts for an atom in state $\ket{n=32,n_{1}=0,n_{2}=n-|m_{l}|-1,m_{l}}$ sitting at the bottom of a lattice well for the one-dimensional, rf-modulated POL described in the previous section (red) and for the TOPOL described in this section (blue). (b) Rabi frequencies coupling a state with $m_{l}$ to $m_{l}+2$ for the one-dimensional, rf-modulated POL (red) and with $m_{l}$ to $m_{l}+1$ for the TOPOL (blue).}
\label{fig:topol}
\end{center}
\end{figure}

 As a result, the electronic ponderomotive potential averaged over a phase-modulation cycle becomes
\begin{multline}
V_{p}(\textbf{r})=\frac{e^{2}}{c\epsilon_{0}m_{e}\omega^{2}_{L}}\bigg[\sqrt{I^{(+)}_{1}I^{(-)}_{1}}J_{0}
(\beta)\\\times\cos{(2kx)}
+\sqrt{I^{(+)}_{2}I^{(-)}_{2}}J_{0}(\beta)\cos{(2ky)}\\
+\frac{I^{(+)}_{1}J^{2}_{0}(\beta)}{2}+\frac{I^{(+)}_{2}J^{2}_{0}(\beta)}{2}+\frac{I^{(-)}_{1}}{2}+\frac{I^{(-)}_{2}}{2}\bigg],
\end{multline}
where $I^{(+)}_{i}(I^{(-)}_{i})$ is the forwards(backwards)-propagating peak intensity of lattice arm $i=1,2$. We ignore the higher-order phase-modulation terms. Numerically determined ponderomotive lattice shifts, $V_{ad,n_{2},m_{l}}$, for atoms situated at the bottom of a well for each state $\ket{m_{l}}$ in the $n=32$ hydrogenic manifold are shown in Fig.~\ref{fig:topol}(a) and compared with the shifts in the previous section for the case of $I_{1}^{(+)}=I_{2}^{(+)}=I_{1}^{(-)}=I_{2}^{(-)}\simeq0.907$ MW/cm$^{2}$ and $J_{0}(\beta)=0.17$. We also set $k=2\pi/532$ nm and assume sufficiently large stabilization fields $F$ and $B$ polarized along $\hat{z}$ in order to prevent state mixing.

The harmonic orbiting of the trap center at $\omega_{rf}$, the resonant angular frequency for the transitions in the hydrogenic manifold, is modeled by the potential
\begin{multline}
   V_{AF}(x,y,t)=\frac{e^{2}}{c\epsilon_{0}m_{e}\omega_{L}^{2}}\bigg\{\bigg[\sqrt{I^{(+)}_{1}I^{(-)}_{1}}J_{1}(\beta)\\
    \times\sin{(2kx)}-i\sqrt{I_{2}^{(+)}I_{2}^{(-)}}J_{1}(\beta)\sin{(2ky)}\bigg]\\
    \times e^{i\omega_{rf}t}+\bigg[\sqrt{I_{1}^{(+)}I_{1}^{(-)}}J_{1}(\beta)\sin{(2kx)}+i\sqrt{I^{(+)}_{2}I^{(-)}_{2}}\\
    \times J_{1}(\beta)\sin{(2ky)}\bigg]e^{-i\omega_{rf}t}\bigg\}.
\end{multline}
Under the assumption that the atom's center-of-mass coincides with a lattice-well minimum and that the atom is small
\begin{multline}
\label{eq:topol_pot}
V_{AF}(\textbf{r}_{e},t)\simeq\frac{\sqrt{8\pi/3}e^{2}kr_{e}}{c\epsilon_{0} m_{e}\omega^{2}_{L}}\bigg\{\bigg[\sqrt{I_{1}^{(+)}I_{1}^{(-)}}J_{1}(\beta_{x})\\
+\sqrt{I_{2}^{(+)}I_{2}^{(-)}}J_{1}(\beta)\bigg]\bigg[Y^{-1}_{1}(\theta_{e},\phi_{e})e^{i\omega_{rf}t}-Y^{1}_{1}(\theta_{e},\phi_{e})e^{-i\omega_{rf}t}\bigg]\\
+\bigg[\sqrt{I_{1}^{(+)}I_{1}^{(-)}}J_{1}(\beta)
-\sqrt{I_{2}^{(+)}I_{2}^{(-)}}J_{1}(\beta)\bigg]\bigg[Y^{-1}_{1}(\theta_{e},\phi_{e})e^{i\omega_{rf}t}\\
-Y^{1}_{1}(\theta_{e},\phi_{e})e^{-i\omega_{rf}t}\bigg]\bigg\}.
\end{multline}
If $\sqrt{I_{1}^{(+)}I_{1}^{(-)}}=\sqrt{I_{2}^{(+)}I_{2}^{(-)}}$, transitions from an unwanted helicity into lower-$|m_{l}|$ states cannot be driven. How well this condition is met determines how well the effective rf field is circularly polarized. After making the rotating-wave approximation and assuming that $\sqrt{I_{1}^{(+)}I_{1}^{(-)}}=\sqrt{I_{2}^{(+)}I_{2}^{(-)}}=\sqrt{I^{(+)}I^{(-)}}$, we arrive at an approximate dressed-atom Rabi frequency from Eq.~\ref{eq:topol_pot}, coupling states $\ket{n,n_{1}=0;m_{l},n_{2}}$ and $\ket{n,n_{1}=0;m_{l}+1,n_{2}-1}$, of
\begin{multline}
\Omega_{m_{l},m_{l}+1}\simeq\frac{4\sqrt{2}e^{2}\sqrt{I^{(+)}I^{(-)}}J_{1}(\beta)k}{\hbar c\epsilon_{0}m_{e}\omega^{2}_{L}}\\
\sum_{ll'}(-1)^{-n-n_{2}+l'+l}(r_{e})^{l'}_{l}
(2l+1)(2l'+1)\\
\times\bigg(\begin{matrix}
     \frac{n-1}{2} & \frac{n-1}{2} & l  \\
     \frac{m_{l}-n_{2}}{2} & \frac{m_{l}+n_{2}}{2} & -m_{l}\\
\end{matrix}\bigg)\bigg(\begin{matrix}
     l & 1 & l' \\
     m_{l} & 1 & -m_{l}-1
\end{matrix}\bigg)\bigg(\begin{matrix}
     l & 1 & l' \\
     0 & 0 & 0
\end{matrix}\bigg)\\
\times
\bigg(\begin{matrix}
     \frac{n-1}{2} & \frac{n-1}{2} & l'  \\
     \frac{m_{l}-n_{2}}{2}+1 & \frac{m_{l}+n_{2}}{2} & -m_{l}\\.
\end{matrix}\bigg).
\end{multline}
We numerically calculate these Rabi frequencies and plot them in Fig.~\ref{fig:topol}(b) for the case of $I^{(+)}=I^{(-)}\simeq0.907$~MW/cm$^{2}$ and $J_{1}(\beta)=0.57$. For comparison, we also display the Rabi frequencies of the one-dimensional optical lattice described in the previous section. \\
\par In addition to the advantage of preventing ``leakage" transitions, the TOPOL clearly provides stronger couplings. This is mainly because the ponderomotive term of the Hamiltonian provides a dipole-like potential which is proportional to a factor of $k$ instead of $k^{2}$ as for the previously-discussed lattice that provides quadrupole-like couplings for atoms at the center of the well. Also, notice that the variation in ponderomotive lattice shifts is lower for the TOPOL because the phase modulation reduces the trap depth by setting $J_{0}(\beta)=0.17$. For a RAP scheme, this would allow more efficient transfer because all dressed states involved would meet closer to degeneracy when the modulation frequency is chirped. By using quantum optimal control theory by varying the TOPOL parameters, one could engineer pulses of $\beta_{x}\cos{(\omega_{rf}t)}$ and $\beta_{y}\sin{(\omega_{rf}t)}$ to obtain fast transfer to the CS on the order of $\sim$ ns and observe coherent Rabi oscillations between the F-state and the CS~\cite{Signoles2017,Patsch2018}.
\section{Discussion}
We now give a discussion comparing each of our three proposed methods with each other and with the traditional methods of circularization that require quasi-static electromagnetic fields and free-space rf radiation. Among the traditional methods of circularization, the RAP method can be separated into two regimes based on the rf-induced coupling strength between atomic states. The method of RAP into the CS with weak rf couplings, initially performed in \cite{Hulet1983}, consists of a series of sequential transitions of $\Delta m_{l}=1$. A highly pure ensemble of CSs at the end of the RAP method is the main benefit of this scheme, in addition to the ease of only requiring linearly polarized rf radiation. However, RAP in the weak-field regime suffers from the extended duration of the process due to the need to meet the adiabaticity condition for each consecutive transition (see Eq.~\ref{eq:diabatic_rate}). Another drawback is that this scheme limits the principal quantum number of the CS to $\sim 60$. Additionally, the long ramping time of the dressed-state eigenenergies in this scheme limits the coherence time of the CS in applications that require a superposition of CSs and other hydrogenic states \cite{Dietsche2019}. For the strong-field regime of RAP, where multiple hydrogenic states are excited at once, the procedure does not require as long ramping times and can be used in conjunction with quantum optimal control theory \cite{Patsch2018,Signoles2017} to minimize the time of passage by means of pulse engineering. The disadvantage of this strong-field regime, however, is the fact that large rf couplings of a linearly polarized rf field will drive ``leakage" transitions that decrease $|m_{l}|$ and make the resulting Rydberg wave packet more elliptical. Therefore, this regime of coupling often requires purely $\sigma^{+}$ or $\sigma^{-}$-polarized rf radiation \cite{Lutwak1997,Signoles2017}, or a sufficiently large magnetic field parallel to the electrostatic field \cite{Nussenzveig193}.

Advantages of the crossed-fields method include the versatility of circularizing within manifolds of higher principal quantum numbers. Also, there is no need for rf radiation at the location of the Rydberg atoms. As mentioned earlier, the crossed-fields method is susceptible to mixing of the CS with low-$|m_{l}|$ states by means of residual electric fields. The crossed-fields method is also not suitable for applications that require a quantization axis defined by the electrostatic field \cite{Ramos2017,Lutwak1997}, as in such applications, one would have to suddenly turn on an electric field that is exactly parallel to the magnetic field. This effects a non-adiabatic transition of the CSs through a multi-level crossing that takes the atoms from the magnetically-stabilized to the electrically-stabilized regime. If there is any remaining perpendicular component of the electric field, the CS becomes contaminated in this process by reduction of $|m_{l}|$.
\par For preparation of an ensemble of cold CSs, laser cooling, trapping, and circularizaiton can all be done with laser fields at appropriate frequencies using the methods proposed in this article based on an additional term in the minimal coupling Hamiltonian. In place of constructing in-vacuum and external electrodes and magnetic coils for the static and rf fields of the traditional methods, there is the technical convenience of aligning laser beams external to the vacuum chamber. We propose three schemes in this article in order to cover a versatile range of experimental contexts from high-precision spectroscopy to long-range interactions of Rydberg atoms.  The method of circularization using LG beams would prepare a dilute, macroscopic sample of CSs that would increase the signal-to-noise ratio in spectroscopic experiments on CSs yet reduce density-dependent line broadening due to the low density of CS atoms produced in this method. Furthermore, because this method affords a single, direct coupling of the F-state to the CS, the rate of circularization is faster than in low-field RAP methods, allowing the preparation of CSs to be done in low-intensity laser fields and weak, static stabilization fields, $F$ and $B$, that are parallel. These weak fields provide smaller perturbations to the Rydberg ensemble, which benefits experiments in high-precision spectroscopy. Circularization via LG laser modes faces the experimental challenge of stabilizing a beat frequency at the order of a few THz and yields a low circularization efficiency of $\sim5\%$. 

The one-dimensional, rf-modulated POL presented in section 3 yields an appealing efficiency of $89\%$, is experimentally simple to construct with an acousto-optic modulator, and provides the control of Rydberg-Rydberg collisions with lattice depths. Additionally, this scheme could be useful for the study of magnetic phase transitions in a one-dimensional chain of Rydberg states \cite{Nguyen2018}. Strong-coupling RAP using this scheme requires the application of a large magnetic field in order to prevent ``leakage" transitions that could be experimentally difficult to switch because of eddy currents; this method also faces the issue of $m_{l}$-dependent ponderomotive shifts that need to be controlled, as exhibited in Fig.~\ref{fig:topol}(a). 

A superior method to the one-dimensional rf-modulated POL is the TOPOL presented in section 4 that does not require a large magnetic field for strong ponderomotive couplings due to the prevention of ``leakage" transitions. With the TOPOL, a two-dimensional sample of CSs can be prepared with a spatial selectivity at the diffraction limit of the lattice beams. Furthermore, the TOPOL provides stronger couplings than the one-dimensional POL, as exhibited in Fig.~\ref{fig:topol}(b), because the time-dependent ponderomotive potential effects an electric-dipole-like coupling of a circularly polarized rf field. One can also implement quantum optimal control theory for this configuration in order to select pulses for the rf modulation of the lattice that transfer the F-states to the CS in the shortest amount of time. This fast transfer minimizes decoherence due to stray electric and magnetic fields and would allow realization of applications in quantum sensing and simulations of two-dimensional Ising models \cite{Dietsche2019,Labuhn2016,Bernien2017}. However, for the TOPOL, failure in matching the intensities and phases of the modulating rf and optical beams in each lattice arm, as well as the polarizations of forward and backwards-propagating lattice beams will result in an effective elliptically polarized rf field from the ponderomotive coupling that would yield a low circularization fidelity. Thus, careful optical alignment is required for the TOPOL.

A key difference between our all-optical methods and the traditional circularization schemes is the requirement of trapping the atoms with optical fields in order to realize the ponderomotive atom-field couplings in the Hamiltonian. We therefore deem our methods not suitable for experiments with hot atomic beams because the atoms must be slow enough to be captured in the optical traps. 

\section{Conclusion}
In summary, we have proposed and discussed three experimental schemes for optical circularization of Rydberg atoms using ponderomotive laser traps. These theoretical investigations demonstrate the versatility of the emerging subfield of ponderomotive interactions with Rydberg atoms. Our proposals could address the difficulty of initializing a quantum system of CSs by using optical couplings instead of static and rf fields. It would be interesting to investigate multilevel Rydberg systems involving the direct optical coupling of a direct, first-order optical coupling of an F-state with a CS~\cite{Han2018}, afforded by the method we proposed in section 2. Additionally, the TOPOL discussed in section 4 would allow a convenient means of initializing a long-lived, two-dimensional Ising model simulator~\cite{Labuhn2016, Bernien2017}. Another proposed quantum simulator that would benefit from these three discussed methods would investigate the angular momentum transport of flexible Rydberg aggregates~\cite{Aliyu2018}. Furthermore, all three of these methods would advance the field of engineering Rydberg wave packets~\cite{Schumacher1995}. 
\section{Acknowledgements}
This work was supported by NSF Grant No. PHY1806809 and NASA Grant No.NNH13ZTT002N NRA.

\bibliographystyle{apsrev4-1}
\bibliography{mybibliography}
\end{document}